\documentclass{article}
\usepackage{arxiv}

\usepackage[utf8]{inputenc} 
\usepackage[T1]{fontenc}    
\usepackage{hyperref}       
\usepackage{url}            
\usepackage{booktabs}       
\usepackage{amsfonts}       
\usepackage{nicefrac}       
\usepackage{microtype}      
\usepackage{lipsum}		
\usepackage{graphicx}
\usepackage{cite}

\usepackage{dsfont}
\usepackage{array}
\usepackage{amssymb, amsmath, amsthm}
\usepackage{graphicx}
\usepackage{lmodern,url}
\usepackage{makecell} 
\usepackage{cancel}
\usepackage{multirow}
\usepackage{microtype}
\usepackage{lineno}
\usepackage{xspace}
\usepackage{xcolor}
\usepackage{siunitx}
\usepackage{todonotes}
\usepackage{booktabs}
\usepackage[ruled,vlined]{algorithm2e}
\usepackage{optidef}
\usepackage{mathdots}
\usepackage{subcaption}
\usepackage{csquotes}
\usepackage{caption}
\newcommand{\pa}{\partial}
\newcommand{\dis}{\displaystyle}

\newcommand{\tabref}[1]{\tablename~\ref{#1}}

\usepackage{amsmath}
\usepackage{tikz}
\usetikzlibrary{fit}
\usetikzlibrary{positioning}
\tikzset{%
  highlight/.style={rectangle,rounded corners,fill=red!15,draw,fill opacity=0.25,thick,inner sep=0pt}
}

\definecolor{mygreen}{RGB}{160, 242,182}

\newcommand{\xunderbrace}[2][\vphantom{\dfrac{A}{A}}]{\underbrace{#1#2}}
\definecolor{mypurple}{RGB}{236, 223, 234}

\allowdisplaybreaks

\newcommand{\Nhatobs}{\hat{N}^{\text{obs}}}

\newcommand{\EH}{E_{i}^{H}}

\newcommand{\ET}{E_{i}^{Q}}

\newcommand{\Ht}{I_{i}^{H}}
\newcommand{\Hta}{I_{i}^{H,a}}
\newcommand{\Hts}{I_{i}^{H,s}}
\newcommand{\Tt}{I_{i}^{Q}}

\newcommand{\SM}{\frac{S}{M}}

\newcommand{\latRate}{\rho}
\newcommand{\Phit}{\Phi_i(t)}

\newcommand{\RelContacts}{k_t}
\newcommand{\RtH}{\RelContacts R_0}

\newcommand{\ROi}{\ensuremath{R_{0}^{i}}\xspace}
\newcommand{\kpoe}{\ensuremath{K_{\rm POE}}\xspace}
\newcommand{\kcom}{\ensuremath{K_{\rm COM}}\xspace}
\newcommand{\fit}{\ensuremath{f_i(t)}\xspace}
\newcommand{\fihat}{\ensuremath{\hat{f}_i}\xspace}
\newcommand{\fihatpoe}{\ensuremath{\hat{f}_i^{\rm POE}}\xspace}
\newcommand{\fihatcom}{\ensuremath{\hat{f}_i^{\rm COM}}\xspace}
\newcommand{\Tin}{\ensuremath{T^{\rm in}}\xspace}
\newcommand{\Tmin}{\ensuremath{\Delta_{\rm time}}\xspace}
\newcommand{\Klim}{\ensuremath{K_{\rm country}^{\rm lim}}\xspace}

\title{Model-based assessment of sampling protocols for infectious disease genomic surveillance}

\usepackage{authblk}

\author[1*]{Sebastian Contreras}
\author[2]{Karen Y. Oróstica}
\author[3]{Anamaria Daza-Sanchez}
\author[1]{Joel Wagner}
\author[1]{Philipp Dönges}
\author[4]{David Medina-Ortiz}
\author[5]{Matias Jara}
\author[2]{Ricardo Verdugo}
\author[3,5]{Carlos Conca}
\author[1,6*]{Viola Priesemann}
\author[4,7*]{Álvaro Olivera-Nappa}

\affil[1]{Max Planck Institute for Dynamics and Self-Organization, Am Fa{\ss}berg 17, 37077 G\"ottingen, Germany.}
\affil[2]{Instituto de Investigación Interdisciplinaria, Vicerrectoría Académica, Universidad de Talca, Talca, Chile.}
\affil[3]{Centre for Biotechnology and Bioengineering, Universidad de Chile, Santiago, Chile.}
\affil[4]{Departamento de Ingenier\'ia en Computaci\'on, Universidad de Magallanes, Punta Arenas, Chile.}
\affil[5]{Departamento de Ingeniería Matemática, Universidad de Chile, Santiago, Chile.}
\affil[6]{Institute for the Dynamics of Complex Systems, University of G\"ottingen, G\"ottingen, Germany.}
\affil[7]{Departamento de Ingeniería Química, Biotecnología y Materiales, Universidad de Chile, Santiago, Chile.}
\affil[ ]{{$*$} Corresponding Authors: Sebastian Contreras (sebastian.contreras@ds.mpg.de), Viola Priesemann (viola.priesemann@ds.mpg.de), and Álvaro Olivera-Nappa (aolivera@ing.uchile.cl)}

\date{}

\renewcommand{\headeright}{ }
\renewcommand{\undertitle}{ }

\begin{document}
\maketitle
\begin{abstract} 
Genomic surveillance of infectious diseases allows monitoring circulating and emerging variants and quantifying their epidemic potential. 
However, due to the high costs associated with genomic sequencing, only a limited number of samples can be analysed. Thus, it is critical to understand how sampling impacts the information generated. 
Here, we combine a compartmental model for the spread of COVID-19 (distinguishing several SARS-CoV-2 variants) with different sampling strategies to assess their impact on genomic surveillance. In particular, we compare \textit{adaptive sampling}, i.e., dynamically reallocating resources between screening at points of entry and inside communities, and \textit{constant sampling}, i.e., assigning fixed resources to the two locations. 
We show that adaptive sampling uncovers new variants up to five weeks earlier than constant sampling, significantly reducing detection delays and estimation errors. This advantage is most prominent at low sequencing rates. Although increasing the sequencing rate has a similar effect, the marginal benefits of doing so may not always justify the associated costs. 
Consequently, it is convenient for countries with comparatively few resources to operate at lower sequencing rates, thereby profiting the most from adaptive sampling. Finally, our methodology can be readily adapted to study undersampling in other dynamical systems.
\end{abstract} 
\clearpage
\section{Introduction}

Genomic sequencing tools help to characterise and keep track of the genetic properties of pathogens causing infectious diseases and strongly contribute to evidence-based decision-making in public health \cite{stark2019use, muellner2016next}. High-throughput, next-generation sequencing technologies (NGS) have substantially reduced sequencing costs over the past 15 years \cite{slatko2018overview}, thereby bringing them closer to routine clinical and public health practices \cite{besser2018next}. An important example is the genomic surveillance of infectious diseases, where the mutational dynamics of a particular pathogen (and variants thereof) are tracked and quantified \cite{armstrong2019pathogen}. In the context of the COVID-19 pandemic, genomic surveillance has unveiled the rapid evolution of SARS-CoV-2 and signalled the emergence of variants with increased transmissibility and partial immune escape (e.g., those labelled as Variants of Concern VoC)  \cite{chen2022global,obermeyer2021analysis,orostica2021mutational,zhu2020novel,GISAID,wu2020new}.

The snapshots provided by genomic surveillance serve three primary purposes \cite{oude2021next,orostica2022new}; i) to signal the introduction of novel variants to a country through surveillance at points of entry (POEs) or detect emerging variants within the communities, ii) to quantify the fraction of the total cases detected in community transmission that these variants caused (thereby enabling the quantification of their spreading rate \cite{orostica2021mutational,obermeyer2021analysis}), and iii) to design and tailor diagnosis and therapeutic alternatives (e.g., drugs and vaccines). However, how reliable this information is depends on i) the quality of the \textit{sampling protocol}, i.e., the strategy to select which PCR-positive samples would be sent for genomic sequencing  \cite{orostica2022new}, and ii) the total number of samples analysed per week (i.e., the \textit{sequencing rate}).

Although official recommendations state that sampling protocols should be coordinated, adaptive, representative, and serve differential purposes \cite{who2021guidance,orostica2022new}, the guidelines to achieve these goals lack a quantitative analysis of the benefits that these concepts bring. Moreover, the optimal strategy is not universal but is expected to depend on a country's resource availability. Despite decreasing costs for NGS, the economic barriers raised by the high equipment and training costs remain prohibitory for low-to-middle income countries \cite{brito2021global,chen2022global,cyranoski_alarming_2021,malick2021genomic,bartlow2021cooperative,helmy2016limited,armstrong2019pathogen}. Therefore, exploring how different sampling protocols for genomic surveillance determine the information we gather can help these nations to optimise resource allocation. 

In our work, we propose a hybrid (deterministic/stochastic) model-based approach to assess the effectiveness of sampling protocols for genomic surveillance on a country-level scale. We focus on answering how to allocate limited sequencing resources best to ensure the early detection of variants, in a setting where: i) sequencing capacity is limited, ii) new variants are imported and enter the system through the POEs, as an external input, and iii) sampling is representative and corrects for potential heterogeneities in the population.  First, we simulate the ground truth dynamics for the simultaneous spread of several SARS-CoV-2 variants using a deterministic differential equations model. Then, we build a stochastic framework to emulate sampling over temporal trends, enabling us to assess the performance of arbitrarily complex sampling protocols. In particular, we compare \textit{adaptive} sampling (dynamically reallocating resources between screening at points of entry and communities according to new variants' detection) and \textit{constant} sampling (sequencing a fixed number of samples from each source). We assess the performance of each strategy through their i) variant detection delay (time between the introduction and first detection of a variant in community transmission), and ii) how well these can approximate the ground truth dynamics by estimating the share of the total cases that each of these represent (and thereby inform inference models). Besides, our approach constitutes a methodological advance that can be readily adapted to model sampling in other systems far from equilibrium. Altogether, we provide new quantitative insights to optimise sampling protocols for genomic surveillance and evidence for the benefits of using adaptive sampling, especially in countries with limited sequencing capacity.

\clearpage
\section{Methods overview}

\subsection{Hybrid approach to simulate genomic surveillance in realistic settings} 

To assess and compare the performance of adaptive and constant sampling protocols, we need to test them under the same conditions. Besides, to determine which one approximates the true underlying dynamics better, we need to (approximately) know the system's state at each time (i.e., its ground truth). As that is not possible in real settings, we propose a model-based hybrid approach: First, we formulate a deterministic mathematical model to represent the simultaneous spread of several SARS-CoV-2 variants in a closed population and thereby produce the ground truth of our system, i.e., the variant-resolved COVID-19 incidence over time. Second, we compare different protocols to determine the origin of samples that will be sent for sequencing (i.e., sampling protocols). Finally, we evaluate the performance of each protocol by quantifying i) how well the share of new COVID-19 cases caused by each variant at a given time is represented and ii) the delay between the true introduction of a new variant and its first detection in community transmission (hereafter \textit{detection delay}). 

In the following, we introduce general aspects of the model for disease spread, the numerical experiments and scenarios that we propose, and the implementation of sampling for genomic surveillance. Full details can be found in Methods, Section~\ref{sec:methods}.

\subsection{Model overview} 

\begin{figure}[!ht]
    \centering
    \includegraphics[width=90mm]{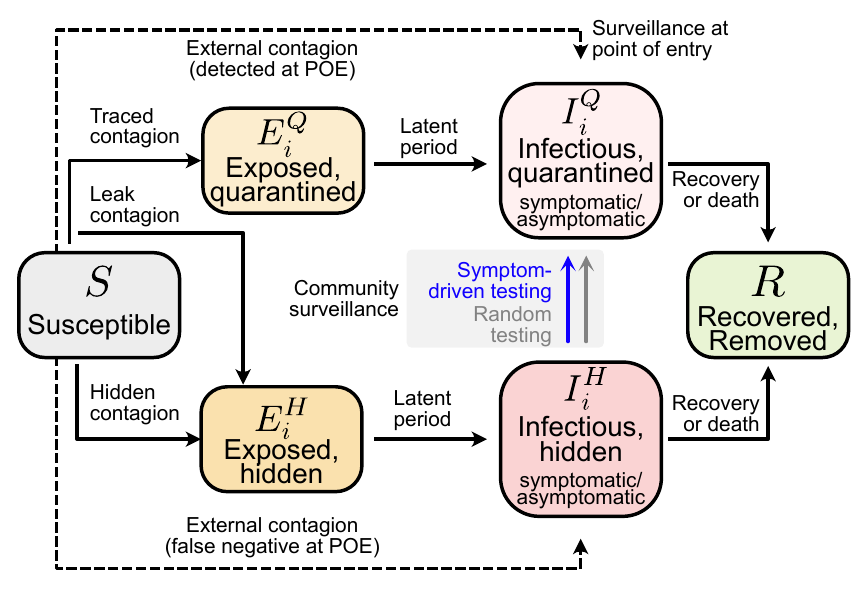}
    \caption{%
        \textbf{Flowchart of the complete model.} The solid blocks in the diagram represent different SEIR compartments for both hidden and quarantined individuals. Hidden cases are further divided into symptomatic and asymptomatic carriers.
        Solid lines represent the natural progression of the infection (contagion, latent period, and recovery). Dashed lines account for the external influx of infections, while testing is represented by arrows moving individuals from the hidden to the quarantined infectious compartment. Quarantined compartments, which contribute less to the spreading of the disease, are coloured with paler shades.
        }
    \label{fig:figure_1}
\end{figure}

We study the spread of COVID-19 using a deterministic ordinary differential equations (ODE) susceptible-exposed-infectious-recovered (SEIR) compartmental model, where several SARS-CoV-2 variants can spread simultaneously. In our model (adapted from~\cite{contreras2021low,contreras2021challenges} and schematised in Fig.~\ref{fig:figure_1}), we distinguish between two contributions of infections: hidden and quarantined. Hidden infections are those where the infector is unaware of being infectious. Therefore, hidden chains propagate unnoticed in the communities until detected via testing. On the other hand, quarantined infectious individuals can also infect others due to imperfect isolation and compliance. However, quarantined infections spread at a much lower rate than hidden infections. We assume that individuals are equally susceptible to all SARS-CoV-2 variants before they had any infection, and after recovery, they obtain cross-immunity against infection. 
Hence, there is not explicit immune-escape in the timeframe considered. 

From the point of view of most (in particular small) countries, new SARS-CoV-2 variants were often introduced from abroad over the course of the COVID-19 pandemic. To reflect that in the model, we include a non-zero influx $\Phit$ of new cases that acquired the virus variant $i$ abroad, reentering our system through points of entry (POEs). These imported variants, labelled as variants of concern (VoCs) abroad, subsequently spread in the communities. In addition, to increase our model's flexibility, we distinguish between symptomatic and asymptomatic infections and allow for potentially different asymptomatic ratios and test sensitivity across variants. Finally, since testing is explicitly considered in our model, we can estimate the "observed" new COVID-19 cases detected via PCR testing (and the collection of samples that can be selected for sequencing). Although we know, by construction, which SARS-CoV-2 variant caused each case, this information is only revealed in each sampling strategy if the sample is selected for sequencing. 

\subsection{Scenarios for the baseline spreading dynamics} 

We formulate different scenarios for the baseline spreading dynamics of COVID-19 at a country-scale, thereby evaluating different patterns for the theoretical waves of incidence. We assume that variants can only be imported from abroad, and that they are introduced to the system through the POEs (and represent this as an external input to the system of ODEs). As a general rule, in each scenario, we set (1) initial conditions, (2) time of introduction of each variant, and (3) for each variant the transmissibility (through their adjusted reproduction number \ROi). Note that the \ROi values capture both the variant base transmissibility and the reductions induced by non-pharmaceutical interventions (NPIs) and hygiene measures. Thus, \ROi are lower than typically reported base reproduction numbers for SARS-CoV-2 variants (Tables~\ref{tab:influx_y_transmisibilidad_experimentos} and~\ref{tab:Parametros}). We initialise our system with a single variant in the population and introduce the following ones as an influx to the system acting at different times, defined per scenario. After solving the system of ODEs that define our model, we estimate the "observed" cases at POEs and communities, and accumulate them to obtain weekly trends (typical temporal resolution for sequencing rates). 

\begin{figure}[!ht]
    \centering
    \includegraphics[width=170mm]{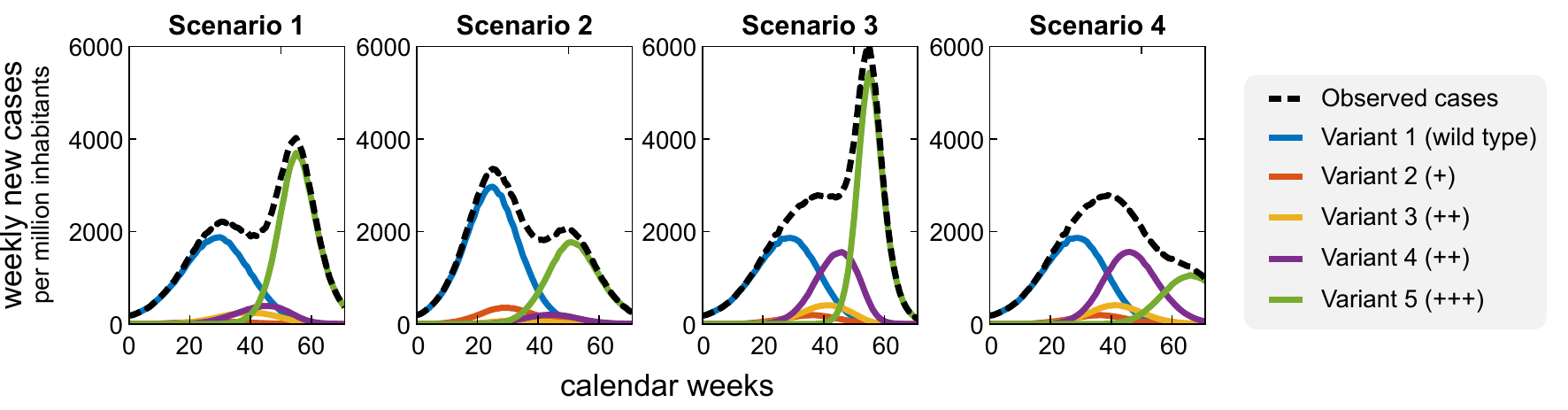}
    \caption{%
        \textbf{Scenarios considered for waves of incidence}. To evaluate sampling strategies, we define a set of scenarios that differ in the transmissibility and time of introduction of a new variant, represented by different colours (variants' spreading parameters are reported in Table~\ref{tab:influx_y_transmisibilidad_experimentos}). Without genomic surveillance, policymakers would only observe the bulk trend of PCR-positive COVID-19 new cases (dashed line) without noticing new variants' emergence and replacement dynamics. In the figure legend, "+" represents increased transmissibility compared to that of the wild type. Scenario 1: double peak (second higher), two dominant variants. Scenario 2: double peak (first higher), two dominant variants. Scenario 3: double peak (second higher), three dominant variants. Scenario 4: single peak, three dominant variants.
        A systematic analysis of wave patterns is provided in Supplementary Materials, Section~S2.
        }
    \label{fig:figure_2}
\end{figure}

We use our model to answer the following question: Given a specific sampling protocol for genomic surveillance (adaptive or constant sampling), how long will it take until we detect a newly introduced variant in community transmission? To that end, we study a system where one variant is dominant, and a second one (with higher transmissibility) enters the system at a given time. We systematically explore different combinations of transmissibility and time of introduction, generating continuous wave patterns. The results are summarised in Supplementary Section~S2. Additionally, we decided to illustrate our methodology by studying four markedly different scenarios, summarised in Fig.~\ref{fig:figure_2} and Table~\ref{tab:influx_y_transmisibilidad_experimentos}. We choose the scenarios in Fig.~\ref{fig:figure_2} motivated by typical wave patterns observed during the COVID-19 pandemic (e.g.,~\cite{contreras2021low}). In particular, scenarios 1 and 2 represent a situation where only two variants drive the wave, and other VoCs do not reach a significant share of the observed cases at any point (e.g., the wild type and Alpha waves in 2020/2021). Scenario 3 shows a situation where the "taking over" of a second variant is replaced by a third, much more transmissible variant (as the emergence of the Omicron VoC in 2021~\cite{orostica2021mutational,orostica2022new}), leading to a markedly higher second peak. The last scenario (4) illustrates the situation where a single peak wave in observed cases hides multiple peaks of different variants. Note that different variants in this context do not refer to a particular VoC, but to a determined configuration of transmissibility and time of introduction, which can vary across scenarios (as described in Table~\ref{tab:influx_y_transmisibilidad_experimentos}).

\subsection{Constant and adaptive sampling protocols} 

\begin{figure}[!ht]
    \centering
    \includegraphics[width=170mm]{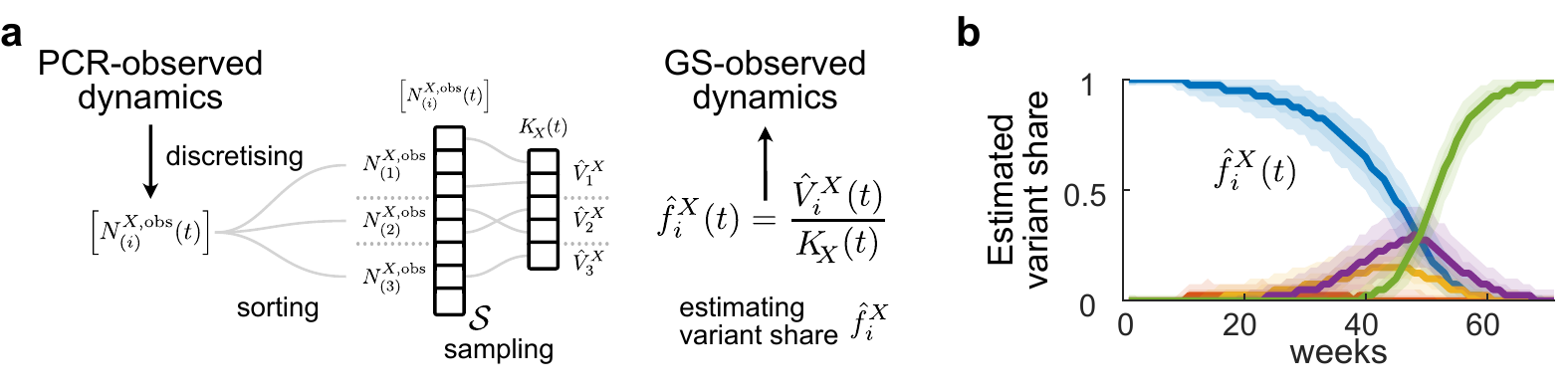}
    \caption{%
        \textbf{Description of the genomic surveillance framework implemented in this work.} \textbf{a:} We discretise the "observed" weekly new COVID-19 cases, understood as a collection of PCR-positive samples eligible for genomic surveillance (GS), and sort them according to the variant that caused them to create a vector $\mathcal{S}$ of labels. We then select $K_X(t)$ ($X:$ point of entry or community) random samples from this pool of tests (i.e., entries of $\mathcal{S}$) and reveal their label. We count the number of times that a variant $i$ was detected this week, $\hat{V}_i(t)$, and estimate the share that they represent of total cases $\hat{f}_i^{X}(t)$, defined as the quotient $\hat{V}_i(t)/K_X(t)$. Finally, we repeat the random selection of samples an arbitrary number of times using different random seeds and study the resulting distributions (example in \textbf{b}).
        }
    \label{fig:figure_3}
\end{figure}

After estimating and discretising the weekly "observed" trends of new COVID-19 cases at POEs and communities, we sort them according to the variant they represent (see Fig.~\ref{fig:figure_3}). Of the resulting vector $\mathcal{S}$ of length $[N_{(i)}^{X,\,{\rm obs}}(t)]$ (with $X \in \{\rm POE, COM\}$) representing the eligible positive samples collected in week $t$, we select $K_X(t)$ entries to be sequenced, thereby revealing their label (i.e., the variant $i$ to which they belong). This is done by drawing $K_X(t)$ random numbers between $1$ and $[N_{(i)}^{X,\,{\rm obs}}(t)]$, and identifying to what variant the corresponding entries in $\mathcal{S}$ belong. We repeat the sampling stage several times to obtain meaningful statistics. As the overall sequencing rate $K$ is constant, we must ensure that $K_{\rm POE}(t)+ K_{\rm COM}(t) = K$.

As described above, we test two alternative sampling protocols for genomic surveillance: i) \textit{constant} sampling, i.e., destining a fixed amount of sequencing capacities to samples collected at POEs and communities, and ii) \textit{adaptive} sampling, i.e.,  dynamically reallocating sequencing capacity between POEs and communities. Reallocation of sequencing capacity in the adaptive protocol is determined by the following criterion: First, as we assume that after detecting a variant at POEs other cases would probably have bypassed the entry screening and thus be already spreading in the population, parts of the sequencing capacity at POEs should be reallocated to surveillance in communities. Second, if the variant the fraction of the total cases represented by a given variant (hereafter, \textit{variant share} $f_i$) estimated from the community contagion (\fihatcom) does not change much over time (i.e., that $|f_i(t)-f_i(t-1)|\leq \varepsilon$ arbitrarily small), the variant replacement dynamics have reached an equilibrium. In that case, the baseline sequencing capacity at POEs should be reinstated. We provide a detailed description of the mathematical formulation of the protocols (e.g., the equilibrium criteria for equilibrium in variant shares in the adaptive case) and parameter values in the Methods section, subsections~\ref{supsec:sampling_protocols} and~\ref{supsec:adaptive_sampling}.

\section{Results}

\subsection{Adaptive sampling protocols significantly reduce variant detection delays and estimation errors}

\begin{figure}[!ht]
    \centering
    \includegraphics[width=170mm]{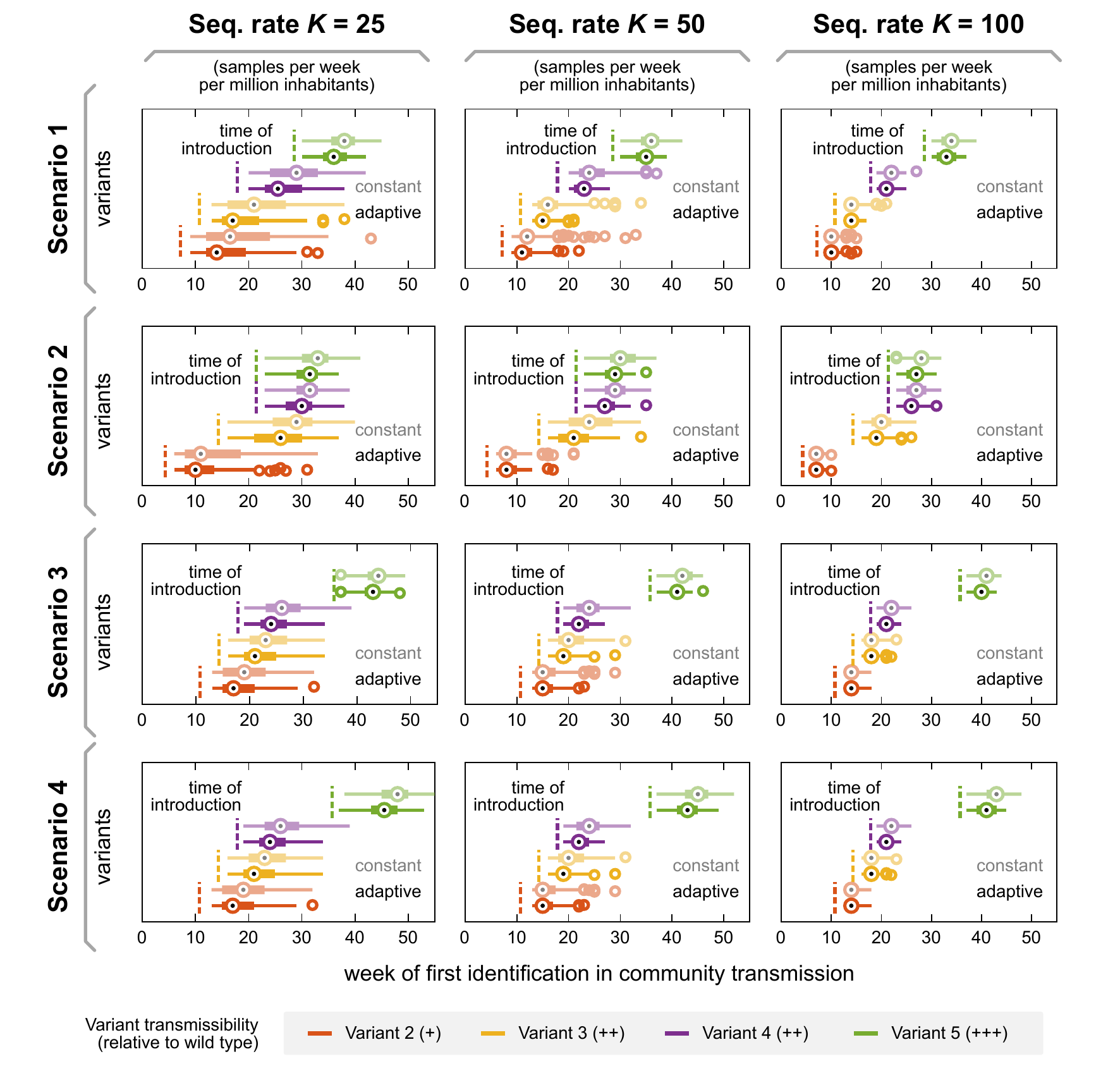}
    \caption{%
        \textbf{Variant detection time across scenarios for different sampling protocols and sequencing rates.} An \textbf{adaptive} sampling protocol for genomic surveillance (i.e., dynamically reallocating sampling resources between POEs and community) reduces the time between variant introduction (dashed line) and the first detection in community transmission significantly when compared to a protocol with \textbf{constant} sampling in POEs and community (solid vs faded, see statistical significance levels in Tabs.~S1 and~S2, and description of both strategies in Section "Constant and adaptive sampling protocols" and Methods). Variants are colour-coded as in Fig.~\ref{fig:figure_2}, and "+" represents their increased transmissibility compared to the wild type. Here, boxplots represent results using different random seeds for the sampling stage ($m=100$ realisations). Besides reducing the expected time of first detection, an adaptive protocol also reduces the variance of the distribution. However, this effect is secondary to increasing the sequencing rate $K$, which can reduce both median times and variability more drastically. Black dots represent medians, boxes the interquartile range of the distribution (upper quartile: 0.75 quantile, lower quartile: 0.25 quantile), and whiskers its range excluding outliers. Outliers (represented as circles) are defined as elements more than 1.5 interquartile ranges above the upper quartile or below the lower quartile. 
        }
    \label{fig:figure_4}
\end{figure}

We assess the efficacy of the sampling protocols described above across scenarios, repeating the random selection of samples $m=100$ times. We find that an adaptive protocol significantly reduces the (expected) detection delay compared to a protocol with constant sampling. Besides, the overall delay distribution in the former is narrower than in the latter. While this result is consistent across scenarios (see Fig.~\ref{fig:figure_4}), the reduction in dispersion achieved through an adaptive protocol is secondary to increasing the sequencing rate $K$.

Depending on the value of $K$, an adaptive protocol can detect a variant in community transmission a couple of weeks earlier than a constant sampling protocol. Although we focus on and emphasise improvements regarding the time of variant detection, an adaptive sampling protocol also improves the accuracy of the estimated trends for variant shares (see Supplementary Material, Section~S4).

\subsection{The marginal benefits of increasing the sequencing rate decline quickly}

\begin{figure}[!ht]
    \centering
    \includegraphics[width=170mm]{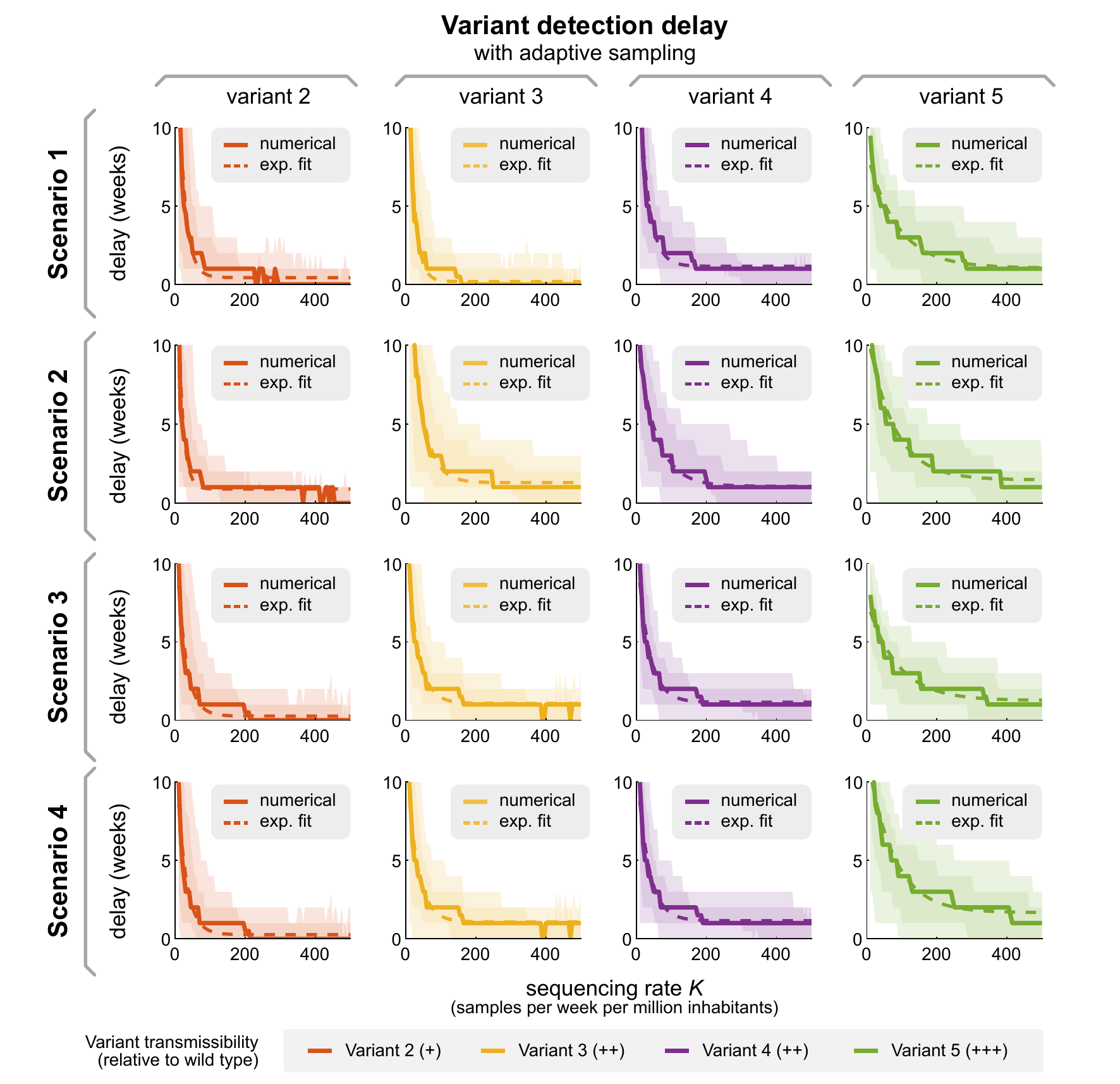}
    \caption{%
        \textbf{Across scenarios, increasing the sequencing rate $K$ strongly decreases the detection delay for all variants.} Solid lines represent the median delay between true introduction and first detection of different SARS-CoV-2 variants across scenarios, and dashed lines represent proposed exponential function (cf. Eq.~\ref{eq:Delay}). Shaded areas denote sample variability (dark: 68\%, light: 95\%).  Results for a constant sampling protocols are provided in Supplementary Fig.~S2.
        }
    \label{fig:figure_5}
\end{figure}

We now analyse the expected detection delay $D$ for both adaptive and constant sampling protocols in all scenarios, as a function of the sequencing rate $K$. We observe that settings with low base sequencing rates would substantially profit from increasing it, by means of reducing the detection delay more steeply when adding an extra unit of $K$. However, such improvement quickly reaches a plateau; further reductions of the expected detection delay would require major increases in $K$ (cf. Fig.~\ref{fig:figure_5}). In other words, these improvements would cost a much higher price.

\begin{figure}[!ht]
    \centering
    \includegraphics[width=170mm]{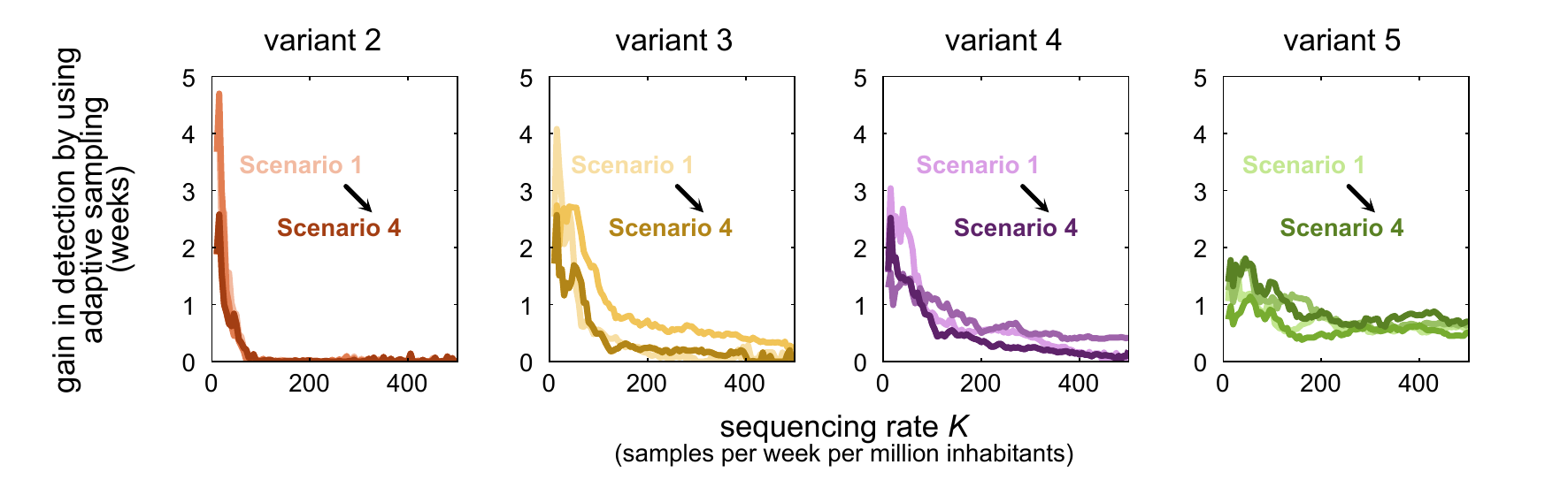}
    \caption{%
        \textbf{Across scenarios, the gains of using an adaptive sampling protocol instead of constant sampling are higher for lower values of $K$.} Although the overall dependency on $K$ is similar for both sampling protocols, i.e., both decay exponentially, differences in the detection delay obtained by each (averaged across realisations) are markedly stronger for lower $K$ in all scenarios. Furthermore, how quickly this difference vanishes when increasing $K$ also seems to depend on the number of variants spreading simultaneously. Note that scenarios 3 and 4 only differ in the parametrisation of the last variant (5) (see Table~\ref{tab:influx_y_transmisibilidad_experimentos}), and thus coincide for variants 2-4.}
    \label{fig:figure_6}
\end{figure}

The sharp decrease in the detection delay $D$ when increasing $K$ resembles an exponential decay. For analytical purposes only, we fit an exponential function to the empirical trends for the median detection delay in Fig.~\ref{fig:figure_5}. The equation for the exponential fit is given by

\begin{equation}\label{eq:Delay}
    D(K) = D_0 \exp\left(-\frac{K-K_{0}}{K_{\rm ref}}\right) + D_{\infty},
\end{equation}

where $D_0$ represents the expected delay for the minimum sampling $K_0$, $K_{\rm ref}$ is a reference sequencing rate, and $D_{\infty}$ represents the minimum delay we can reach. Fitted trends agree well with our simulations for both adaptive and constant sampling protocols (cf. the overlap between solid and dashed lines in Figs.~\ref{fig:figure_5} and~S2). We also performed this experiment for the constant sampling protocol (see Supplementary Fig.~S2). While both graphics look very alike, there are marked differences between the two. For example, differences in the average detection delay for both protocols can be as large as five weeks when the sequencing rate $K$ is low and quickly decline as we increase the sequencing rate $K$ (Fig.~\ref{fig:figure_6}). However, the benefits of using an adaptive protocol also depend on the number of variants co-circulating; if there are several variants in the pool of infections, improvements by using an adaptive sampling persist to higher sequencing rates (Fig.~\ref{fig:figure_6}).

In the following section, we use the analytical approximation we propose for the expected variant detection delay to generalise our results to an economic perspective.

\subsection{Economic assessment of strategies for genomic surveillance}

As the early detection of variants in community transmission allows policymakers to timely implement measures to mitigate their impact, reducing the detection delay would benefit all actors in society. In economic terms, this defines an \textit{utility} function $U(D)$ that increases as we reduce $D$. As per the observations in the previous section, we know that $D$ decreases when we increase the number of samples analysed $K$. The question is then how much extra benefit an extra unit of $K$ would produce, i.e., what is the \textit{marginal utility} of increasing $K$. Speaking against increasing $K$, the \textit{marginal costs} of increasing it should grow linearly while well below the sequencing capacity limit given by country-specific infrastructure ($K_{\rm country}^{\rm lim}$) and should strongly increase when approaching it. Assuming that the variant detection delay profiles remain unchanged across countries, we can study the optimal number of sequences that should be analysed per week. In other words, the sequencing rate from which the marginal benefits reached by increasing $K$ would not justify the required costs. 

\begin{figure}[!ht]
\hspace*{-1 cm}
    \centering
    \includegraphics[width=170mm]{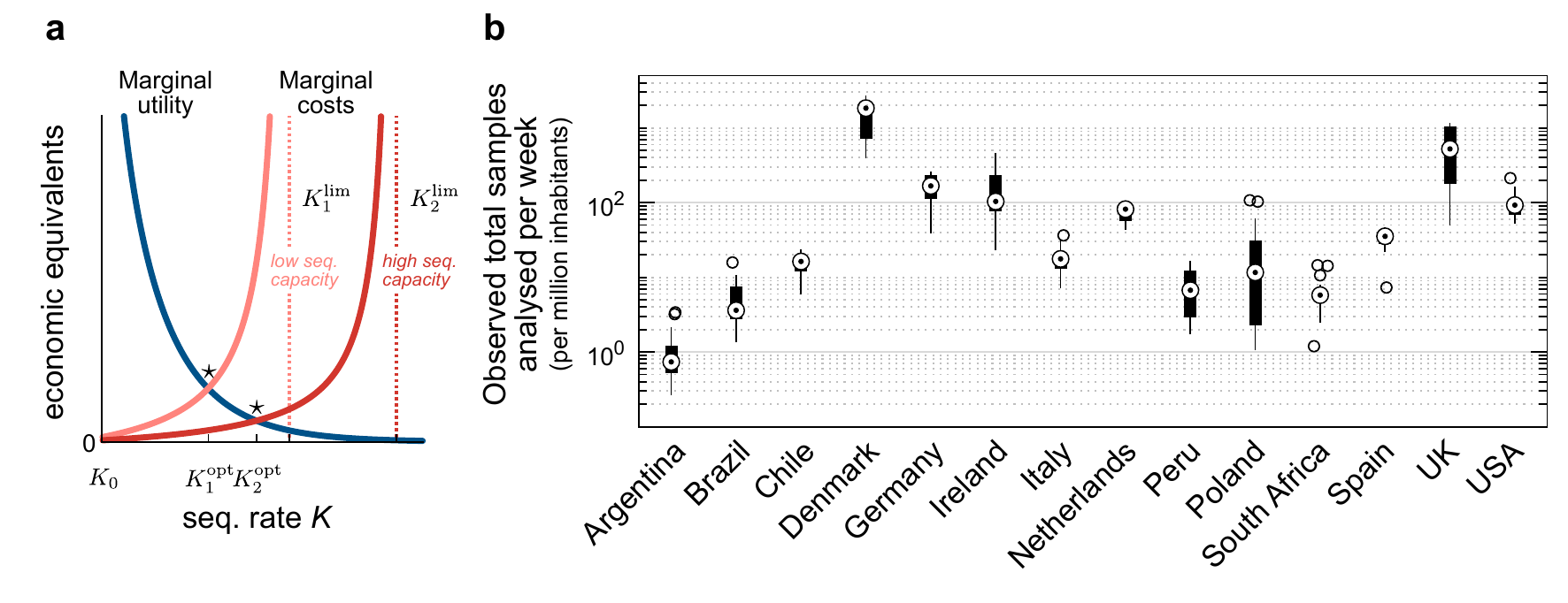}
    \caption{%
        \textbf{Cost-benefit analysis of increasing the sequencing rate in countries with different sequencing capacity.} \textbf{a:} Based on economic terms, countries with less installed sequencing capacity $K^{\rm lim}$ will see their operational cost escalate considerably at a lower number of sequences analysed per week ($K$), finding their operational optima at lower values. \textbf{b:} Observed weekly sequencing rate $K^{\rm obs}$ for different countries worldwide, normalised per million inhabitants. The logarithmic scale used to represent the y axis facilitates comparing observed sequencing rates across countries, where differences can be of orders of magnitude. Boxplots describe the $K^{\rm obs}$ values observed between Feb. 2nd, and Jun. 4th, 2022.}
    \label{fig:figure_7}
\end{figure}

Using the mathematical formulation for the marginal utility and costs presented in Supplementary Section~S1 (Eqs.~2 and~3), we schematise the criteria for economically-optimal sequencing in two types of countries (Fig.~\ref{fig:figure_7}a). On the one hand, countries with high installed sequencing capacities \Klim can increment the number of samples they analyse per week without incurring in higher additional costs. On the other hand, countries with less sequencing infrastructure or specialised workforce will see their costs increase disproportionally larger for lower $K$, finding their sequencing optimum $K_{1}^{\rm lim}$ at fewer samples per week. In that case, instead of searching to increase $K$ beyond  $K_{1}^{\rm lim}$, these countries would find it more rewarding to reallocate those resources into other active interventions, such as subsidies for lockdowns and the distribution of hygiene materials to the general population. While these economic principles are clear, how much improvements in reducing $D$ are valued in a given country needs to be quantified in economic terms by local policymakers.

Analysing real-world data of the observed sequencing rates $K$ in countries worldwide, we see a sizeable week-to-week variability (see Fig.~\ref{fig:figure_7}b). For example, countries in the global north have higher observed sequencing rates and dispersion overall (week-to-week variations in $K$). This can be due to the protocols they follow for sequencing; rather than being the installed capacity \Klim which limits the rate, they aim to send a fixed fraction of the observed new cases for sequencing \cite{brito_global_2021}. 

\section{Discussion}

In our manuscript, we used a hybrid model-based approach combining deterministic ODE models with a stochastic sampling framework to assess the effectiveness of different sampling protocols for genomic surveillance. Our quantitative insights support the benefits of using adaptive sampling, where sequencing efforts are reallocated between surveillance at points of entry (POEs) and communities according to the progression of the disease. We showed that adaptive sampling protocols outperform protocols where a constant amount of sequencing capacity is allocated to POE and community samples. These results hold across incidence wave patterns (scenarios and systematic analysis, cf.~Supplementary Section~S2) and values for the sequencing rate $K$. 

Compared with a constant sampling protocol, adaptive sampling can reduce the expected detection delay of introduced variants by a couple of weeks at the same sequencing rate $K$, especially when operating at low sequencing rates. Timely detecting a new variant is critical to mitigating its potential impacts, especially for diseases that spread fast. For example, considering the doubling time of Omicron VoC infections (between 1.5 and 3 days in its initial stages \cite{torjesenn2021omicron_doublingtime}), detecting its introduction two weeks later (i.e., $\sim 5$ doubling times later) implies dealing with an incidence $\sim$30x larger. Therefore, using an adaptive protocol allows policymakers to react earlier to emerging public health threats, thereby facilitating containment through test-trace-and-isolate \cite{contreras2021low, contreras2021challenges} and minimising disruptions to everyday life and economies \cite{czypionka2022benefits,prati2021psychological}. We also showed that $K$ is the strongest determinant for reducing the detection delay $D$. In fact, $D$ declines exponentially when increasing $K$. However, this also implies that the benefits earned by expanding $K$ by an extra unit decline similarly and would soon not justify the high costs incurred. Thus, there is a cost-effective optimal $K$, which depends on the \textit{installed} sequencing capacity of a given country \Klim and how local policymakers weigh reductions in $D$ in economic terms. 

Despite decreasing sequencing costs induced by technological advancements, genomic surveillance is still costly as it requires specialised equipment, high-performance computing capability,  and specialist personnel. Thus, it raises economic barriers that not all countries can circumvent \cite{brito2021global, chen2022global,cyranoski_alarming_2021,malick2021genomic,bartlow2021cooperative,helmy2016limited,armstrong2019pathogen}. This translates to little elasticity to changing the sequencing rate $K$, and inspired our assumption of setting it constant. For example, in Chile, despite the governmental and private investments in genomic surveillance, the sequencing rate is around 400 samples per week (i.e., ~20 samples per million inhabs.), at least two orders of magnitude lower than the UK's surveillance program, with a sequencing rate larger than 10~\% of their positive weekly tests (\cite{chen2022global,covid2020integrated,brito_global_2021}, and Fig.~\ref{fig:figure_7}b). Therefore, most decisions have been based on trends of the current reproduction number, which, however, does not capture the spreading dynamics of VoCs \cite{contreras2020statistically,contreras2020real,medina2020country,freire2021heterogeneous,Sanchez-Daza2022}. The situation is similar in other countries in Latin America and the global south, where countries have not reached a sequencing rate of 1~\% of their positive tests \cite{brito_global_2021}. Overall, sequencing rates and genomic surveillance programmes are markedly different between high and low-income countries, where the sequences reported by the former are 12 times higher than the latter. Furthermore, the ratio of confirmed cases sequenced by high-income countries is 16 times higher than that of low-income countries (4.36\% and 0.27\%, respectively) \cite{chen2022global}. This again highlights the economic determinants of success in pandemic control \cite{contreras2022rethinking, smith2020covid, petherick2021worldwide, mena2021socioeconomic, snowden2021covid}.

Following the same line of thought, we formulate our model assuming that the sequencing capacity is the limiting factor for surveillance and apply it to study how different sampling protocols (i.e., distributions of this capacity between points of entry and communities) can help to reduce the variant detection delay. This restricts our analysis to settings where variants enter the system through defined points, and sampling protocols guarantee representativeness and correct for potential heterogeneities among the population. Examples of that are small countries with singular points of entry or isolated communities within a country \cite{orostica2021mutational,gonzalez2021mutations_in_magallanes}. An equally important question in settings with more sequencing capacity is how much sequencing is required to correctly estimate the share of the total cases corresponding to a given variant. This question is thoroughly studied in \cite{wohl2022sampling}, where the authors determine, for different scales, the required number of sequences to be analysed to estimate the share of cases belonging to each variant. This question is critical to determine whether a newly detected variant is taking over and should be considered a VoC. However, as we study a setting resembling a small country where variants are introduced from abroad, it falls outside the scope of this paper.

Our deterministic model for disease spread has certain limitations. For example, we do not include age structure in our model, as we do not intend to quantify the impacts in morbidity/mortality that different variants may cause. We also do not include vaccination or waning immunity, as these are unessential within the time frame we analyse. In fact, in the timeframe we study here, more than 90\% of the population had only one infection, and only a tiny fraction had two or more.
Furthermore, although vaccination effectively reduces morbidity and mortality rates from COVID-19, the effect that this process generates on the transmissibility of a given variant, especially Omicron, is relatively small (e.g., \cite{woodbridge2022viral, kissler2021viral,singanayagam2021community}). Therefore, vaccination was not considered an essential variable in this modelling.
There is also no spatial resolution in the model (as in, e.g., \cite{contreras2020multigroup,freire2021heterogeneous}) as we assume sampling to be representative, and these would only affect wave patterns (which do not compromise our results). Besides, we excluded contact tracing from our model; samples collected within the same infection chain are likely caused by the same SARS-CoV-2 variant (thus, including them would induce selection biases in our analyses). Finally, for simplicity, we assume that genomic surveillance does not affect variant transmissibility while, in real settings, information about a new variant is likely to trigger new interventions. However, this is straightforward to incorporate by including a feedback loop between the estimated variant share $\fihat$ and the overall spreading rate $\beta$ (as, e.g., in~\cite{contreras2021winter,doenges2022interplay,bauer2021relaxing}), and does not play a role in the metrics we analyse (i.e., detection delay and mismatch between estimations and ground truth). Nonetheless, the model is simple enough to serve our objective fully: To produce quantitative insights on the performance of different sampling protocols for genomic surveillance in detecting introduced variants and reducing the uncertainty of their inference with the available resources.

Similar sampling approaches to ours can, in principle, be applied to many other physical problems. Limited sampling poses a challenge when trying to access properties of various complex dynamical systems \cite{levina2017subsampling,wilting2018inferring}. Furthermore, undersampling may introduce systematic bias to observations that need to be corrected \cite{priesemann2009subsampling,wilting2019criticality,levina2022}. This can happen, for example, when assessing \textit{collective} properties, like graph structures in a network or activity clusters spanning large fractions of the system. In the case of detecting SARS-CoV-2 variants, such undersampling bias is not expected because the random selection of a PCR-positive sample is representative. Here, the core challenge is economical; how much we can sample depends on the resources destined for genomic surveillance. Thus, it is crucial to implement methods to maximise the information gathered with the available resources. Our work demonstrates the benefits of using adaptive sampling in genomic surveillance and quantifies the improvements reached by increasing the installed sequencing capacity to reduce the detection delay of newly introduced variants. Besides, the proposed methodology can readily be adapted to study other dynamical systems far from equilibrium or arbitrarily complex sampling protocols. This is crucial to assess current protocols and design contingency plans for current and future global health emergencies, especially in settings where resources are limited.

\section{Methods}\label{sec:methods}

\subsection{Spreading Dynamics}

We propose a modified SEIR-type model to adequately capture COVID-19 spread, where infected individuals can be either symptomatic or asymptomatic, and their infection can be caused by several co-circulating SARS-CoV-2 variants. They belong to hidden ($\EH,\Ht$) or quarantined ($\ET,\Tt$) pools of infections, thus creating in total one compartment of susceptible individuals ($S$), two compartments of exposed individuals ($\EH$, $\ET$), three compartments of infectious individuals ($\Hts$, $\Hta$, $\Tt$), and one compartment for recovered/removed individuals ($R$). Model compartments, transitions between them, and testing mechanisms are illustrated in Fig.~\ref{fig:figure_1}.

New infections are asymptomatic with a variant-specific ratio $\xi_i$, the remaining infections are symptomatic. In all compartments, individuals are removed at a rate $\gamma$ because of recovery or death  (see~\tabref{tab:Parametros} for all parameters). In the hidden pools, the disease spreads according to the population's contact patterns and the base transmissibility of the variants. Here, we parameterise the spreading rate of SARS-CoV-2 variants through their NPI-corrected reproduction number \ROi. In this parameter, we combine the base spreading properties of the variant (as per their base reproduction number) with typical levels of contact reductions induced by moderate restrictions. This reproduction number \ROi reflects the disease spread in the general population without the testing-induced isolation of individuals, nor current immunity levels. Additionally, the hidden pool receives a mobility-induced influx $\Phit$ of new infections. Cases are removed from the hidden pool (i) when detected by testing and put into the quarantined pool $\Tt$, or (ii) due to recovery or death. 

The quarantined exposed and infectious pools $(\ET,\Tt)$ contain those infected individuals who have been tested positive as well as their positively tested contacts. Infectious individuals in $\Tt$ are (imperfectly) isolated; we assume their contacts have been reduced to a fraction $(\nu+\epsilon)$ of the ones they had in pre-COVID-19 times, of which only $\nu$ are captured by the tracing efforts of the health authorities. As traced cases generated by isolated individuals would be of the same SARS-CoV-2 variant, we do not include them into the pool of tests potentially sent for sequencing. The remaining fraction of produced infections, $\epsilon$,  are missed and act as an influx to the hidden pools ($\EH$). Therefore, the overall reproduction number in the $\Tt$ pool is $\left(\nu+\epsilon\right) R_0$. 

\subsection{Testing strategies}

We consider symptom-driven testing and random testing: 

\textbf{Symptom-driven testing} is here defined as applying tests to individuals presenting symptoms of COVID-19. In this context, it is important to note that non-infected individuals can have symptoms similar to those of COVID-19, as many symptoms are rather unspecific. Although symptom-driven testing suffers less from imperfect specificity, it can only uncover symptomatic cases that are willing to be tested (see below). Here, \textit{symptomatic, infectious individuals} are transferred from the hidden to the traced pool at rate $\lambda_{s}$.

\textbf{Random testing} is here defined as applying tests to individuals irrespective of their symptom status or whether they belong to the contact chain of other infected individuals. In our model, random testing transfers infected individuals from the hidden to the quarantined infectious pools with a fixed rate $\lambda_{r}$, irrespective of whether or not they are showing symptoms. 

\subsection{Model Equations}

\begin{align}
\frac{d S}{dt} & =  -\xunderbrace{\gamma\SM\dis\sum_{i=1}^{k}\ROi I_{i}^{H}}_{\text{hidden contagion}} - \xunderbrace{\gamma\SM\dis\sum_{i=1}^{k}(\epsilon+\nu) \ROi I_{i}^{Q}}_{\text{quarantined contagion}} - \xunderbrace{\SM\sum_{i=1}^{k}\Phit}_{\text{ext. influx}} , \label{eq:dSdt}\\
\frac{d \ET}{dt} & =  \xunderbrace{\gamma\SM\nu \ROi I_{i}^{Q}}_{\text{observed contagion}} -\xunderbrace{\latRate \ET}_{\text{end of latency}}, \label{eq:dETdt}\\
\frac{d \EH}{dt} & =  \xunderbrace{\gamma\SM \ROi I_{i}^{H}}_{\text{hidden contagion}}+\xunderbrace{\gamma\SM\epsilon \ROi I_{i}^{Q}}_{\text{leak contagion}}- \xunderbrace{\latRate \EH}_{\text{end of latency}}, \label{eq:dEHdt}\\
\frac{d \Tt}{dt} & =  \xunderbrace{\latRate \ET-\gamma \Tt}_{\text{spreading dynamics}} + \xunderbrace{(\lambda_s+\lambda_r)\eta_i \Hts}_{\text{testing, symptomatic}} + \xunderbrace{\lambda_r \eta_i \Hta}_{\text{testing}} + \xunderbrace{\eta_i\SM\Phit}_{\text{ext. influx (POE detected)}}  ,\label{eq:dIOdt}\\
\frac{d \Hts}{dt} & =  \xunderbrace{\xi_i\latRate \EH-\gamma \Hts}_{\text{spreading dynamics}} - \xunderbrace{(\lambda_s+\lambda_r) \eta_i \Hts}_{\text{testing}} + \xunderbrace{\xi_i(1\!-\!\eta_i)\SM\Phit}_{\text{ext. influx (false negative at POE)}} , \label{eq:dIHsdt}\\
\frac{d \Hta}{dt} & =  \xunderbrace{(1\!-\!\xi_i)\latRate \EH-\gamma \Hta}_{\text{spreading dynamics}} - \xunderbrace{\lambda_r \eta_i \Hta }_{\text{testing}} + \xunderbrace{(1\!-\!\xi_i)(1\!-\!\eta_i)\SM\Phit}_{\text{ext. influx (false negative at POE)}}, \label{eq:dIHadt}\\
\frac{d R}{dt}    & =  \xunderbrace{\gamma\sum_{i=1}^{k} \left(\Tt+\Hta+\Hts\right)}_{\text{recovered/removed individuals}}.   \label{eq:dRdt}
\end{align}

\subsection{Initial conditions}

Let $x$ be the vector collecting the variables of all different pools:

\begin{equation}
    x = [S,\,\ET,\,\EH,\,\Tt,\,\Hts,\,\Hta,\,R].
\end{equation}

We assume a population size of $M=10^6$ individuals, such that $\sum_{i} x_i = M$. We initialise all scenarios with only one variant (the wild type), and the following settings: $\ET(0) = 50$, $\EH(0) = 1050$, $\Tt(0) = 100$, $\Hts(0) = 250$, $\Hta(0) = 750$, (for $i=1$), and $S(0) = 997800$.

\subsection{Modelling the influx of infections and the introduction of new variants}

In our model, we incorporate a mechanism for externally acquired infections, i.e., individuals belonging to the population, but acquiring the virus (and variants thereof) overseas. Explicitly, they appear as an influx $\Phit$, which we model as the overlap of different gamma-distributed pulses and constant contributions of "old" variants. Mathematically, the influx (as a vector of size $k$) is given by

\begin{equation}
    \Phi(t) = \sum_{i=1}^{k} e_i \Phi^{\rm max}_{i}\Gamma(a_i,b_i)(t) + \Phi_{\rm base} \frac{\sum_{i=1}^{k} e_i\mathds{1}\left( \Tin_i + \Tmin \leq t \leq \Tin_{i+1} + 2\Tmin\right)}{1+\sum_{i>1}^{k}\mathds{1}\left(\Tin_i + \Tmin \leq t \leq \Tin_{i+1} + 2\Tmin\right)},
\end{equation}

where $e_i$ are canonical unit vectors, $a_i$ and $b_i$ shape and scale parameters for the Gamma distribution, $\Phi_{\rm base}$ the baseline influx of infections, $\Tin_i$ the time of introduction of the $i$'th variant to the system, and \Tmin represents the time window where an old variant continues appearing in the influx. Values for variant-specific parameters across scenarios are given in Table~\ref{tab:influx_y_transmisibilidad_experimentos}.


\begin{table}\caption{Influx parameters and differential transmissibility of (theoretical) SARS-CoV-2 variants across scenarios.}
\label{tab:influx_y_transmisibilidad_experimentos}
\centering
\begin{tabular}{ccccccc}\toprule
Variant     & Scenario & $\Phi^{\rm max}_{i}$           & $a_i$                & $b_i$                 & \Tin [days]                & $R_0^{i}$ \\\midrule
\multirow{4}{*}{1} & 1          & \multirow{4}{*}{1000} & \multirow{4}{*}{3} & \multirow{4}{*}{4} & \multirow{4}{*}{-5} & 1.5   \\
                   & 2          &                       &                    &                    &                     & 1.6   \\
                   & 3          &                       &                    &                    &                     & 1.5   \\
                   & 4          &                       &                    &                    &                     & 1.5   \\\midrule
\multirow{4}{*}{2} & 1          & \multirow{4}{*}{1500} & \multirow{4}{*}{4} & \multirow{4}{*}{5} & 50                  & 1.5   \\
                   & 2          &                       &                    &                    & 30                  & 1.8   \\
                   & 3          &                       &                    &                    & 75                  & 1.75  \\
                   & 4          &                       &                    &                    & 75                  & 1.75  \\\midrule
\multirow{4}{*}{3} & 1          & \multirow{4}{*}{1000} & \multirow{4}{*}{3} & \multirow{4}{*}{6} & 75                  & 1.8   \\
                   & 2          &                       &                    &                    & 100                 & 2     \\
                   & 3          &                       &                    &                    & 100                 & 2     \\
                   & 4          &                       &                    &                    & 100                 & 2     \\\midrule
\multirow{4}{*}{4} & 1          & \multirow{4}{*}{1500} & \multirow{4}{*}{3} & \multirow{4}{*}{5} & 125                 & 2     \\
                   & 2          &                       &                    &                    & 150                 & 2.5   \\
                   & 3          &                       &                    &                    & 125                 & 2.25  \\
                   & 4          &                       &                    &                    & 125                 & 2.25  \\\midrule
\multirow{4}{*}{5} & 1          & \multirow{4}{*}{500}  & \multirow{4}{*}{2} & \multirow{4}{*}{5} & 200                 & 3     \\
                   & 2          &                       &                    &                    & 150                 & 3     \\
                   & 3          &                       &                    &                    & 250                 & 4.5   \\
                   & 4          &                       &                    &                    & 250                 & 3.5  \\\bottomrule
\end{tabular}
\end{table}

\subsection{Central epidemiological parameters that can be observed}

In the real world, disease spread can only be observed through testing and contact tracing. While the \textit{true} number of daily infections $N$ is a sum of all new infections in the hidden and traced pools, the \textit{observed} number of daily infections $\Nhatobs$ is the number of new infections discovered by testing, tracing, and monitoring of the contacts of those individuals in the quarantined infectious pool $\Tt$, delayed by a variable reporting time. This includes internal contributions as well as contributions from testing and tracing:

\begin{align}
    N_i &= \xunderbrace{\gamma k_t\ROi\SM\Ht}_{\text{hidden contagion}} + \xunderbrace{\gamma \left(\nu+\epsilon\right)\ROi\SM\Tt}_{\text{observed contagion}}+ \xunderbrace{\SM\Phit}_{\text{ext. influx}}
    \label{eq:N}\\
    \hat{N}_{{\rm com}, i}^{\rm obs} &=  \Big[\xunderbrace{\lambda_s \Hts}_{\text{sympt. test}}+ \xunderbrace{\lambda_r \Ht}_{\text{rand. test}}\,\Big]  \circledast \mathcal{K},
    \label{eq:Nreportcom}\\
    \hat{N}_{{\rm POE}, i}^{\rm obs} &=  \Big[\xunderbrace{\eta_i \Phit}_{\text{test at POE}}\,\Big]  \circledast \mathcal{K},
    \label{eq:Nreportpoe}
\end{align}
where $\circledast$ denotes a convolution and $\mathcal{K}$ an empirical probability mass function that models a variable reporting delay, inferred from German data (as the RKI reports the date the test is performed, the delay until the appearance in the database can be inferred): The total delay between testing and reporting a test corresponds to one day more than the expected time the laboratory takes for obtaining results, which is defined as follows: from testing, \SI{50}{\%} of the samples would be reported the next day, \SI{30}{\%} the second day, \SI{10}{\%} the third day, and further delays complete the remaining \SI{10}{\%}, which for simplicity we will truncate at day four. Considering the extra day needed for reporting, the probability mass function for days 0 to 5 would be given by $\mathcal{K}=[0,\,0,\,0.5,\,0.3,\,0.1,\,0.1]$. 

\subsection{Modelling sampling protocols}\label{supsec:sampling_protocols}

As described earlier, we compare two sampling protocols for genomic surveillance throughout the manuscript, constant sampling and adaptive sampling. For the first case, we assume that the number of samples collected at POEs and communities remains constant so that $\kpoe$ and $\kcom$ are constant. In contrast, an adaptive sampling protocol prioritises samples collected at POEs or communities depending on the genomic surveillance findings of the previous weeks. While markedly different, both start from the same baseline $\kpoe(t=0) = \left\lfloor0.6\,K\right\rfloor$, and have the same thresholds $K_{\rm COM}^{\rm min}= 0.6\,K$ and $K_{\rm COM}^{\rm max}= \left\lfloor0.95\,K\right\rfloor$ (although these are meaningful only for the adaptive case). In the following section, we describe the adaptive sampling protocol in detail.

\begin{table}[htp]\caption{Model parameters.}
\label{tab:Parametros}
\centering
\begin{tabular}{l p{5cm} lll p{2cm}}\toprule
Parameter       & Meaning                       & \makecell[l]{Value \\ (default)}    & \makecell[l]{Range\\ }         & Units             &   Source  \\\midrule
$M$             & Population size               & $\num{1000000}$       &               & people          &   -       \\
$R_0^i$           & NPI-corrected reproduction number variant $i$     & 4                  & see Table~\ref{tab:influx_y_transmisibilidad_experimentos}       & \SI{}{-}  & \cite{ZHAO2020214,davies_estimated_2021,torjesenn2021omicron_doublingtime}\\
$\nu$           & Registered contacts (quarantined)& 0.075      &        & \SI{}{-}          &  \cite{contreras2021low}   \\
$\epsilon$      & Lost contacts (quarantined)     & 0.05      & & \SI{}{-}          &  \cite{contreras2021low}    \\
$\gamma$        & Recovery/removal rate         & 0.10      & 0.08--0.12& \SI{}{day^{-1}}  &  \cite{he2020temporal,pan2020time}        \\
$\xi_i$         & Asymptomatic ratio for variant $i$            & 0.32      & 0.15--0.43     & \SI{}{-}          &   \cite{lavezzo2020suppression,wilhelm2021omicron_reduced} \\
$\lambda_s$     & Symptom-driven testing rate   & 0.25      & 0--1 & \SI{}{day^{-1}}          & Assumed \\
$\lambda_r$     & Random testing rate & 0.0      & 0.0--0.1 & \SI{}{day^{-1}}          & Assumed \\
$\eta_i$          & Test sensitivity to variant $i$            & 0.9      &      & \SI{}{-}          &  Assumed  \\
$\Phi_i$         & External influx of variant $i$               & -          & 0--10 & \SI{}{cases\, day^{-1}}        &  Assumed  \\
$\latRate$      & Exposed-to-infectious rate    & 0.25     &           & \SI{}{day^{-1}}       &   \cite{bar2020science,li2020substantial}\\\midrule
$s_\alpha$      & Stiffness adaptive response ($\alpha$)    & 5     &           & \SI{}{-}       &   Assumed      \\
$s_\zeta$      & Stiffness adaptive response ($\zeta$)    & 1     &           & \SI{}{-}       &   Assumed      \\
$\Lambda_{1/2}$      & Middle point sigmoidal response ($\alpha$)    & 0.25     &           & \SI{}{-}       &   Assumed      \\
$\Theta_{1/2}$      &  Middle point sigmoidal response ($\zeta$)  & 10     &           & \SI{}{-}       &   Assumed      \\
$\alpha_0$      & Scaling factor ($\alpha$)   & 5     &           & \SI{}{-}       &   Assumed      \\
$\Delta K_{\rm max}$      &  Max adaptation of community surveillance & 10     &           & \SI{}{-}       &   Assumed      \\
$K_{\rm COM}^{\rm max/min}$      &  Max/min value for community surveillance & -     & 60-95\% $K_{\rm com}^{\rm base}$          & \SI{}{-}       &   Assumed      \\\bottomrule
\end{tabular}%
\end{table}

\begin{table}[htp]\caption{Model variables.}
\label{tab:Variables}
\centering
\begin{tabular}{l p{4cm} l  p{7.5cm} }\toprule
Variable & Meaning & Units & Explanation\\\midrule
$S$ & Susceptible pool & \SI{}{people} & non-infected people that may acquire the virus.  \\
$\ET$ & Exposed pool (quarantined)   & \SI{}{people} & Total quarantined exposed people. \\
$\EH$ & Exposed pool (hidden)    & \SI{}{people} & Total non-traced, non-quarantined exposed people.\\
$\Hts$ & Infectious pool (hidden, symptomatic) & \SI{}{people} & Non-traced, non-quarantined people who are symptomatic.\\
$\Hta$ & Infectious pool (hidden, asymptomatic) & \SI{}{people} & Non-traced, non-quarantined people who are asymptomatic.\\
$\Ht$ & Infectious pool (hidden) & \SI{}{people} & Total non-traced, non-quarantined infectious people. $\Ht = \Hts+\Hta$.\\
$\Tt$ & Infectious pool (quarantined) & \SI{}{people} & Total quarantined infectious people.\\
$N_i$ & New infections (Total, variant $i$) & \SI{}{cases\, day^{-1}} & Given by:  $N = \gamma\RtH\Ht +\gamma \left(\nu+\epsilon\right) \ROi\Tt+\SM\Phit$.\\
$\hat{N}^{\rm obs}_{(i)}$ & Observed new infections (influx to traced pool, variant $i$) &  \SI{}{cases\, day^{-1}} & Daily new cases, observed from the quarantined pool; delayed because of imperfect reporting. \\
$\kpoe$ & Sequencing rate for samples collected at POEs &  \SI{}{samples\, week^{-1}} & Number of POE samples sequenced per million inhabitants and per week. \\
$\kcom$ & Sequencing rate for samples collected from community contagion &  \SI{}{samples\, week^{-1}} & Number of community samples sequenced per million inhabitants and per week. \\
$D$ & Variant detection delay &  \SI{}{weeks} & Time between variant introduction and first detection. \\
\bottomrule
\end{tabular}%
\end{table}

\subsection{Adaptive sampling strategy for genomic surveillance}\label{supsec:adaptive_sampling}

As described previously and in \cite{orostica2022new}, genomic surveillance serves two objectives depending on where samples were collected. On the one hand, if samples are collected at POEs, these signal the introduction of novel variants to the country, and provide an alert of what we should look for in community transmission. On the other hand, samples collected from community transmission provide information on the local features of the spread of such variants, their mutational signatures, and their reproduction numbers. Let $K$ be the total amount of samples that can be sequenced per week (i.e., the sequencing rate), and $\kcom(t)$ and $\kpoe(t)$ be the amount of these that were taken from community contagion and at POEs, respectively, at time $t$. Then, $K = \kcom(t) + \kpoe(t)$. In the adaptive sampling, we allow  $\kcom(t)$ and $\kpoe(t)$ to change over time depending on our estimations for the variant share at POEs and within the community, $\fihatpoe$ and $\fihatcom$. Thus, we define two quantities that will help us decide when to reallocate resources:

\begin{align}
    \Lambda(t) & = \max_{i}\left\{\fihatpoe(t-1) - \fihatcom(t-1)\right\},\\
    \Theta(t) & = \max_{i}\left\{\frac{\pa \fihatcom}{\pa t} (t-1)\right\},
\end{align}

where $\pa$ denotes a discrete derivative. When $\Lambda(t)$ is large, variants being introduced to the country are not yet markedly spreading in the community. When $\Theta(t)$ is large, the replacement dynamics are far from equilibrium --- in either way, we require more sequencing in the community. We use a logistic function to smooth the response, and two auxiliary variables:
\begin{align}
    \alpha(\Lambda) & = \alpha_0\frac{\exp\left(s_\alpha(\Lambda-\Lambda_{1/2})\right)}{1+\exp\left(s_\alpha(\Lambda-\Lambda_{1/2})\right)},\\
    \zeta(\Theta) & = \frac{1}{1+\exp\left(s_\zeta(\Theta-\Theta_{1/2})\right)}.
\end{align}

so that

\begin{equation}
    \frac{\pa \kcom}{\pa t} = \left[\left(\alpha\left(\Lambda\right)- \zeta\left(\Theta\right)\right)\Delta\! k_{\rm max}\right].
\end{equation}

However, as \kcom is bounded between a minimum and maximum value, the effective correction could be lowered to ensure that the following inequality holds

\begin{equation}
    \kcom^{\rm min} \leq \kcom(t) \leq \kcom^{\rm max}.
\end{equation}

\section*{Author Contributions}

Conceptualisation: SC, KYO, AS-D, CC, VP, AO-N\\
Methodology: SC, PD\\
Software: SC, DM-O, MJ\\
Validation: SC, MJ, KYO\\
Formal analysis: SC\\
Investigation: SC, KYO, AS-D\\	
Writing - Original Draft: SC, KYO, AS-D\\	
Writing - Review \& Editing: SC, KYO, AS-D, JW, PD, RV, VP\\
Visualisation: SC\\	
Supervision: CC, VP, AO-N\\	

\section*{Competing Interests}
VP is currently active in various groups to advise the government. 

\section*{Acknowledgements} 
We thank the Priesemann group for exciting discussions and for their valuable input. 

\section*{Funding:} 
SC, JW, PD, and VP received support from the Max-Planck-Society (Max-Planck-Gesellschaft MPRG-Priesemann). SC, JW, PD, and VP received funding by the German Federal Ministry for Education and Research for the RESPINOW project (031L0298). KYO and RV funding by ANID Chile, projects Rapid-Covid COVID0961 and Anillo ACT210085. AS-D, CC, and AO-N acknowledge funding by the Centre for Biotechnology and Bioengineering --- CeBiB (PIA project FB0001, ANID, Chile). DM-O received support from Universidad de Magallanes (MAG-2095 project). 

\subsection*{Data and Materials Availability}
Numerical experiments and data analysis of ODEs was performed using MATLAB R2021a and the DMAKit Python library \cite{medina2020dmakit}. Code and datasets generated are available in the following github repository:\\ \url{https://github.com/Priesemann-Group/sampling_for_genomic_surveillance}

\newpage
\renewcommand{\thefigure}{S\arabic{figure}}
\renewcommand{\figurename}{Supplementary~Figure}
\setcounter{figure}{0}
\renewcommand{\thetable}{S\arabic{table}}
\renewcommand{\tablename}{Supplementary~Table}

\setcounter{table}{0}
\renewcommand{\theequation}{\arabic{equation}}
\setcounter{equation}{0}
\renewcommand{\thesection}{S\arabic{section}}
\setcounter{section}{0}
\setcounter{page}{1}
\section*{Supplementary Material}

\section{Economic assessment of strategies: Calculations}

As an early detection of variants inside the community would allow policymakers to timely enact response measures, reducing the detection delay would be beneficial to all actors. In economical terms, we define an \textit{utility} function, which increases as we reduce $D$:

\begin{equation}\label{eq:utility}
    U(D) = \alpha  \left(D_0-D\right)^{\beta},
\end{equation}

where $\alpha$ and $\beta$ are positive numbers. As per the observations in the previous section, we know that $D$ decreases when we increase the sequencing rate $K$. The question is then how much benefit an extra unit of $K$ would yield. The differential increment in social utility when increasing $K$ is given by $\frac{\pa U}{\pa K}$:

\begin{align}
    \frac{\pa U}{\pa K} & = \alpha\beta\left(D0-D\right)^{\beta -1} \left(-\frac{\pa D}{\pa K}\right),\nonumber\\
        & = \alpha\beta\left(D0-D\right)^{\beta -1} \frac{D_0}{K_{\rm ref}}\exp\left(-\frac{K-K_{0}}{K_{\rm ref}}\right)\label{eq:marginal_utility}
\end{align}

On the other hand, marginal costs should increase linearly while well below the sequencing capacity limit given by country-specific infrastructure ($K_{\rm country}^{\rm lim}$), and should diverge when approaching it. Without incurring in too much detail, we can approximate this curve by

\begin{equation}\label{eq:marginal_costs}
    MC_{\rm country} = \frac{K\eta_{\rm country}}{K_{\rm country}^{\rm lim}-K},
\end{equation}

where $\eta_{\rm country}$ accounts for the cost-efficiency of the operation within  the country. After converting the expressions for marginal utility and costs to economic equivalents, the value for $K$ where these curves meet yields the desirable operation point.

\subsection{Statistical analyses}

We assess whether the detection delay distributions for each sampling protocol were different using the non-parametric Mann-Whitney U test, using the equality of both distributions as the null hypotheses. We implement these tests using the statistical modules of the DMAKit-Lib library \cite{medina2020dmakit}. Finally, we incorporate a statistical significance analysis using the following criteria: i) p-value greater than 0.05, not significant (NS), ii) p-value between 0.01 and 0.05, one level (*), iii) p-value between 0.001 and 0.001, two levels (**), iv) p-value between 0.001 and 0.0001, three levels (***), v) lower p-value at 0.0001, four levels (****).

When comparing the distributions applying the U test, we find that distributions are significantly different for the most transmissible variants and that medians and variances of those obtained for the adaptive sampling are lower than those obtained for the constant sampling (cf. Tab~\ref{suptab:Fig_1_utest}). We validate these results with the Student's t-test, which yield similar results (cf. Tab~\ref{suptab:Fig_1_ttest}), demonstrating the differentiation between distributions of different sampling strategies.

\begin{table}[htp]\caption{Statistical analysis (u-test) results of Fig.~\ref{fig:figure_4}.}
\label{suptab:Fig_1_utest}
\centering
\begin{tabular}{cclclclc}\toprule
                   &         & \multicolumn{2}{c}{Seq. rate K = 25} & \multicolumn{2}{c}{Seq. rate K = 50} & \multicolumn{2}{c}{Seq. rate K = 100} \\\midrule
Experimet          & Variant & p-value   & sig. level    & p-value     & sig. level  & p-value   & sig. level   \\\midrule
\multirow{4}{*}{1} & 1       & 0.037495        & *                  & 0.630882         & NS                & 0.891484         & NS                 \\
                   & 2       & 0.001141        & **                 & 0.001395         & **                & 0.017965         & *                  \\
                   & 3       & 0.00123         & **                 & 4.44E-05         & ****              & 7.64E-05         & ****               \\
                   & 4       & 0.00047         & ***                & 0.001795         & **                & 0.002843         & **                 \\\midrule
\multirow{4}{*}{2} & 1       & 0.071341        & NS                 & 0.617184         & NS                & 0.878771         & NS                 \\
                   & 2       & 0.000462        & ***                & 0.000111         & ***               & 0.002048         & **                 \\
                   & 3       & 0.006206        & **                 & 0.000799         & ***               & 0.000344         & ***                \\
                   & 4       & 0.003654        & **                 & 2.22E-05         & ****              & 0.000106         & ***                \\\midrule
\multirow{4}{*}{3} & 1       & 0.177229        & NS                 & 0.406758         & NS                & 0.969485         & NS                 \\
                   & 2       & 0.006697        & **                 & 0.000459         & ***               & 0.030898         & *                  \\
                   & 3       & 0.002044        & **                 & 0.000294         & ***               & 0.000629         & ***                \\
                   & 4       & 0.006587        & **                 & 5.50E-05         & ****              & 6.47E-05         & ****               \\\midrule
\multirow{4}{*}{4} & 1       & 0.177229        & NS                 & 0.406758         & NS                & 0.969485         & NS                 \\
                   & 2       & 0.006697        & **                 & 0.000459         & ***               & 0.030898         & *                  \\
                   & 3       & 0.002044        & **                 & 0.000294         & ***               & 0.000629         & ***                \\
                   & 4       & 0.000702        & ***                & 0.000112         & ***               & 4.57E-05         & ****               \\\bottomrule
\end{tabular}
\end{table}

\begin{table}[htp]\caption{Statistical analysis (t-test) results of Fig.~\ref{fig:figure_4}.}
\label{suptab:Fig_1_ttest}
\centering
\begin{tabular}{cclclclc}\toprule
                   &         & \multicolumn{2}{c}{Seq. rate K = 25} & \multicolumn{2}{c}{Seq. rate K = 50} & \multicolumn{2}{c}{Seq. rate K = 100} \\\midrule
Experimet          & Variant & p-value   & sig. level   & p-value     & sig. level  & p-value   & sig. level \\\midrule
\multirow{4}{*}{1} & 1       & 0.006655        & **                 & 0.113783         & NS                & 0.865881         & NS                 \\
                   & 2       & 0.00079         & ***                & 0.000146         & ***               & 0.002981         & **                 \\
                   & 3       & 0.000438        & ***                & 1.16E-06         & ****              & 3.20E-05         & ****               \\
                   & 4       & 0.000599        & ***                & 0.001326         & **                & 0.001888         & **                 \\\midrule
\multirow{4}{*}{2} & 1       & 0.008322        & **                 & 0.2005           & NS                & 0.869346         & NS                 \\
                   & 2       & 0.000619        & ***                & 1.28E-05         & ****              & 0.000185         & ***                \\
                   & 3       & 0.006934        & **                 & 0.00034          & ***               & 0.000141         & ***                \\
                   & 4       & 0.005692        & **                 & 2.32E-05         & ****              & 0.000129         & ***                \\\midrule
\multirow{4}{*}{3} & 1       & 0.11882         & NS                 & 0.102448         & NS                & 0.95136          & NS                 \\
                   & 2       & 0.005974        & **                 & 5.29E-05         & ****              & 0.018645         & *                  \\
                   & 3       & 0.000846        & ***                & 2.78E-05         & ****              & 0.000419         & ***                \\
                   & 4       & 0.011327        & *                  & 8.95E-05         & ****              & 6.13E-05         & ****               \\\midrule
\multirow{4}{*}{4} & 1       & 0.11882         & NS                 & 0.102448         & NS                & 0.95136          & NS                 \\
                   & 2       & 0.005974        & **                 & 5.29E-05         & ****              & 0.018645         & *                  \\
                   & 3       & 0.000846        & ***                & 2.78E-05         & ****              & 0.000419         & ***                \\
                   & 4       & 0.000622        & ***                & 6.96E-05         & ****              & 2.18E-05         & ****               \\\bottomrule
\end{tabular}
\end{table}

\clearpage
\section{Complimentary analyses: Systematic analysis of wave patterns}\label{supsec:systematic_analysis}

In this section, we generalise the scenarios presented in the main text by systematically exploring how variant properties affect the variant detection delay. Consider the situation where only one variant spreads in the population and let \ROi be its NPI-adjusted reproduction number. We then introduce a second variant and explore how its reproduction number (relative to the first one) and time of introduction impact the detection delay. We repeat this experiment for different spreading rates and both sampling protocols.

\begin{figure}[!ht]
    \centering
    \includegraphics[width=170mm]{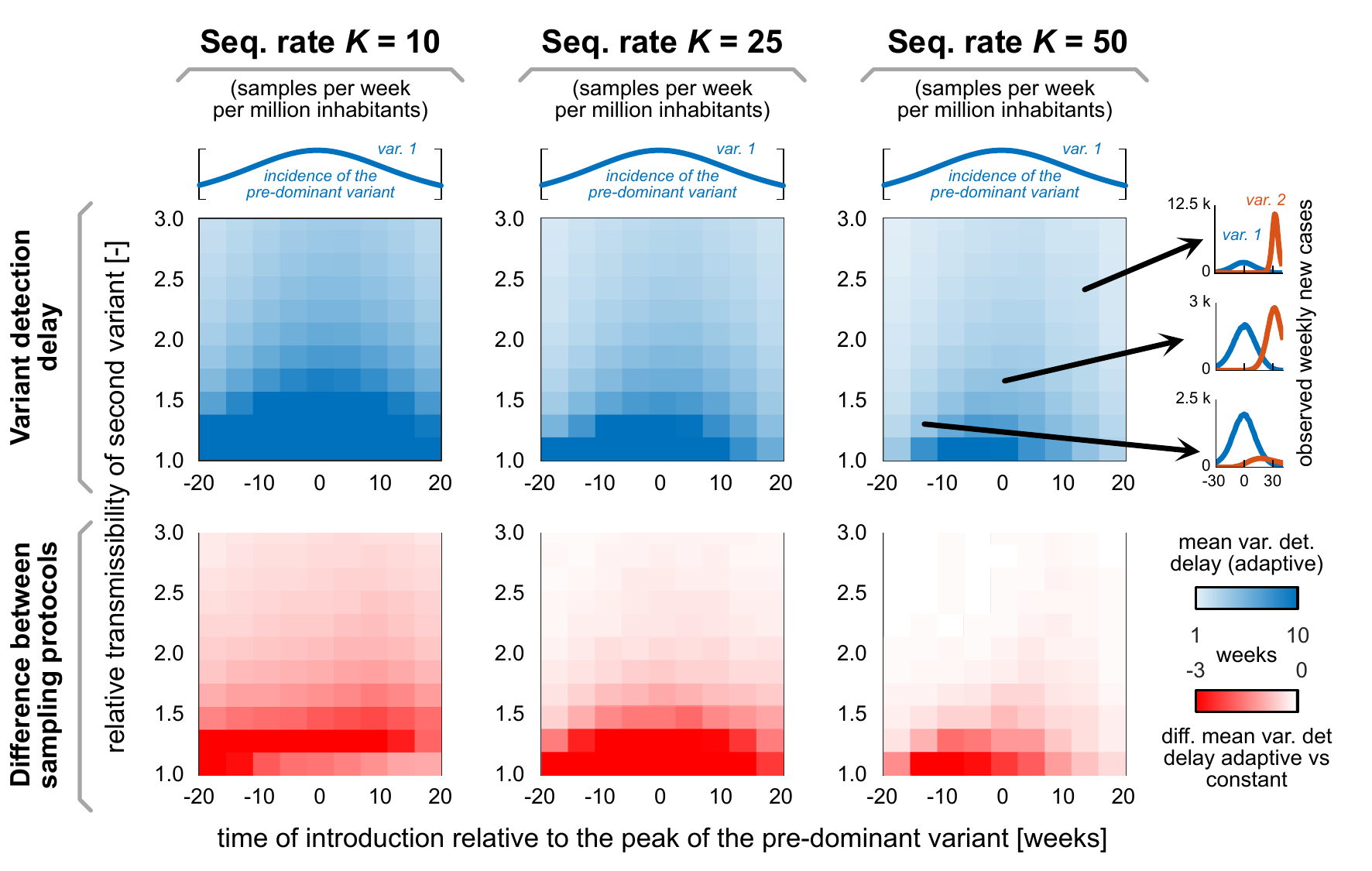}
    \caption{%
        \textbf{Systematic analysis of the influence of variant transmissibility and introduction time on its detection time.} In the main text, we demonstrated that adaptive sampling reduces variant detection delay (compared with constant sampling) for all wave pattern scenarios. Here, we continuously vary the relative transmissibility and introduction time of a variant in a 2-variant system to generalise our results. Consistent with what is shown in the main text, we found that increasing the sequencing rate $K$ reduces the variant detection delay (first row of results). Furthermore, adaptive sampling helps detecting this new variant in community transmission earlier than with constant sampling (second row of results). Averages are calculated over $m=250$ realisations of the numerical experiment.
        }
    \label{supfig:systematic_analysis}
\end{figure}

Using adaptive sampling, the variant detection delay (first row of results in Fig.~\ref{supfig:systematic_analysis}) increases when the time of introduction of the second variant is close to the peak incidence of the first one. This follows frequentist probability rules; at the wave's peak, we have a large pool of tests that will belong to the dominant variant. If that is the moment when we introduce one case of the new variant, detecting it is less likely than when the overall incidence is lower (as we drift from the peak in Fig.~\ref{supfig:systematic_analysis}). As reported in Fig.~\ref{supfig:Results_3}, adaptive sampling outperforms constant sampling especially at low sequencing rates (see the column for $K=10$ in Fig.~\ref{supfig:systematic_analysis}). Besides, an adaptive protocol is also helpful when the incidence of the original variant is high. Altogether, an adaptive protocol consistently outperforms constant sampling for genomic surveillance.

\clearpage
\section{Complimentary analyses: Mirror figures}
\begin{figure}[!ht]
    \centering
    \includegraphics[width=170mm]{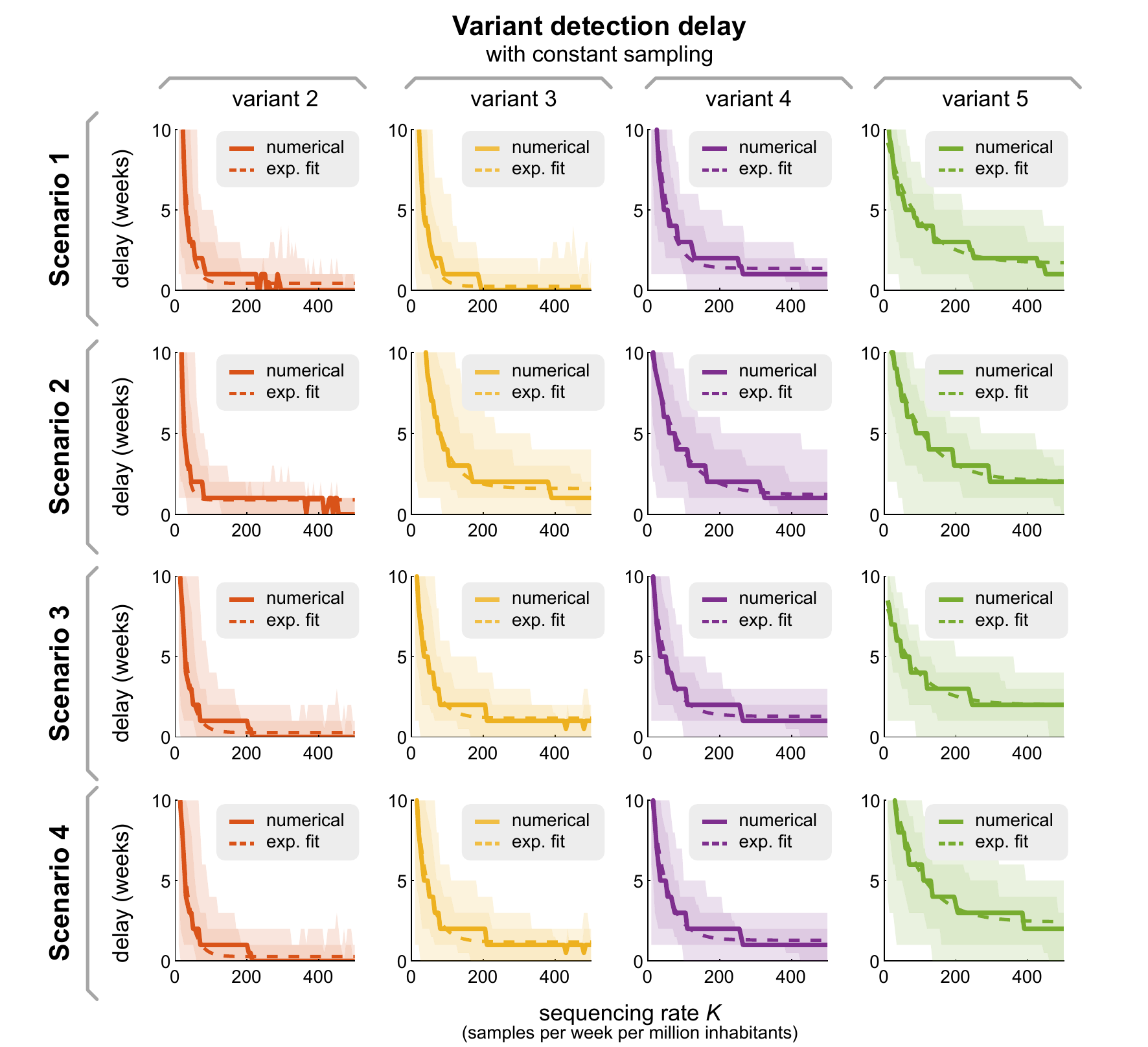}
    \caption{%
        \textbf{Mirror figure of Fig.~\ref{fig:figure_5} when using a constant sampling protocol instead.} Solid lines represent the median delay between true introduction and first detection of different SARS-CoV-2 variants across scenarios, and dashed lines represent proposed exponential function (cf. Eq.~\ref{eq:Delay}). Shaded areas denote sample variability (dark: 68\%, light: 95\%).  
        }
    \label{supfig:Results_2}
\end{figure}

\section{Complimentary analyses: variability reduction}\label{supsec:abs_error}

\begin{figure}[!ht]
    \centering
    \includegraphics[width=170mm]{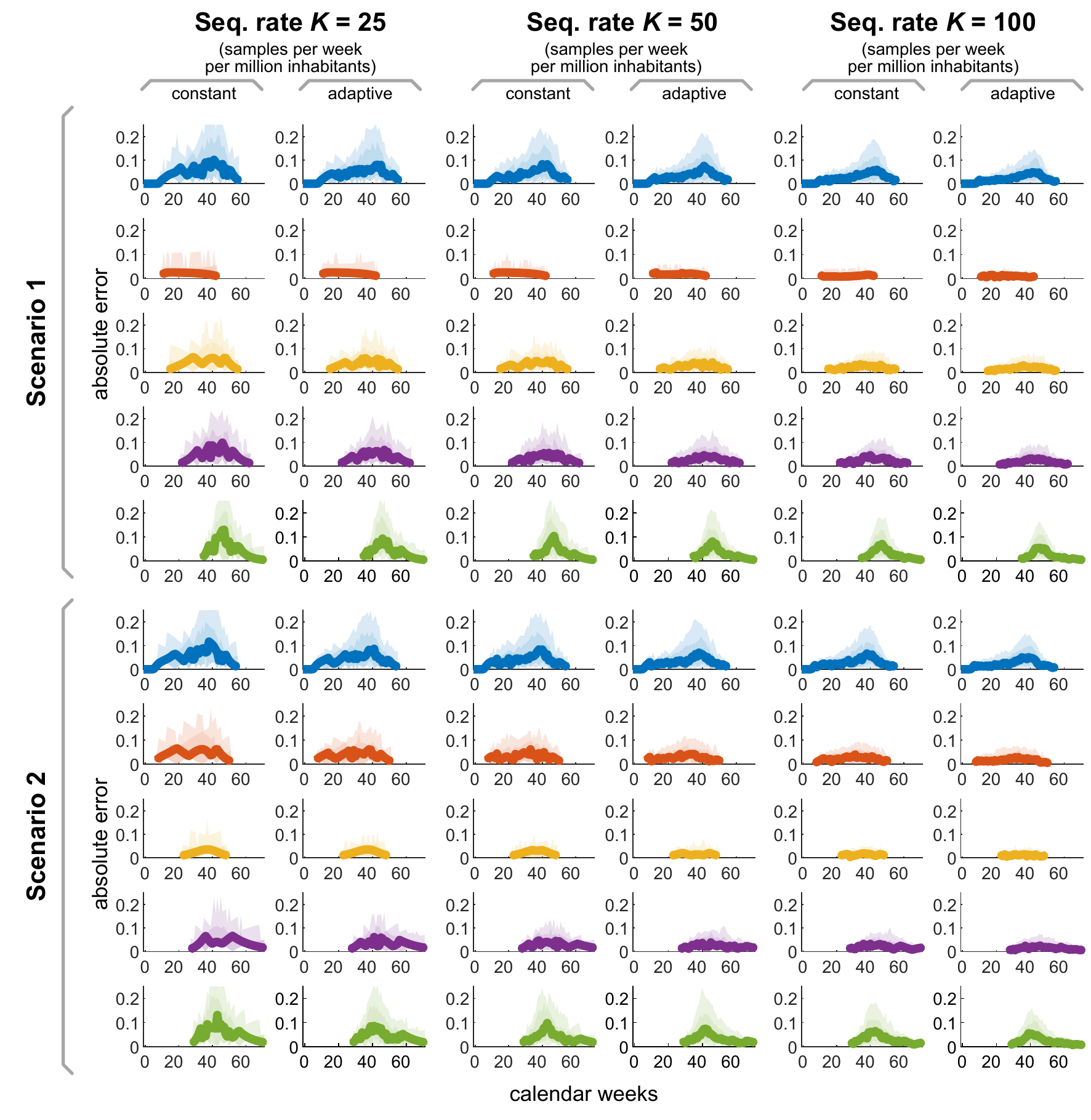}
    \caption{%
        \textbf{Absolute error trends for the variant shares in scenarios 1 and 2.} Solid lines represent the median absolute error between the estimated variant share $\fihat(t)$ and its ground truth \fit. Shaded areas denote sample variability (dark: 68\%, light: 95\%).
        }
    \label{supfig:Results_3}
\end{figure}

\begin{figure}[!ht]
    \centering
    \includegraphics[width=170mm]{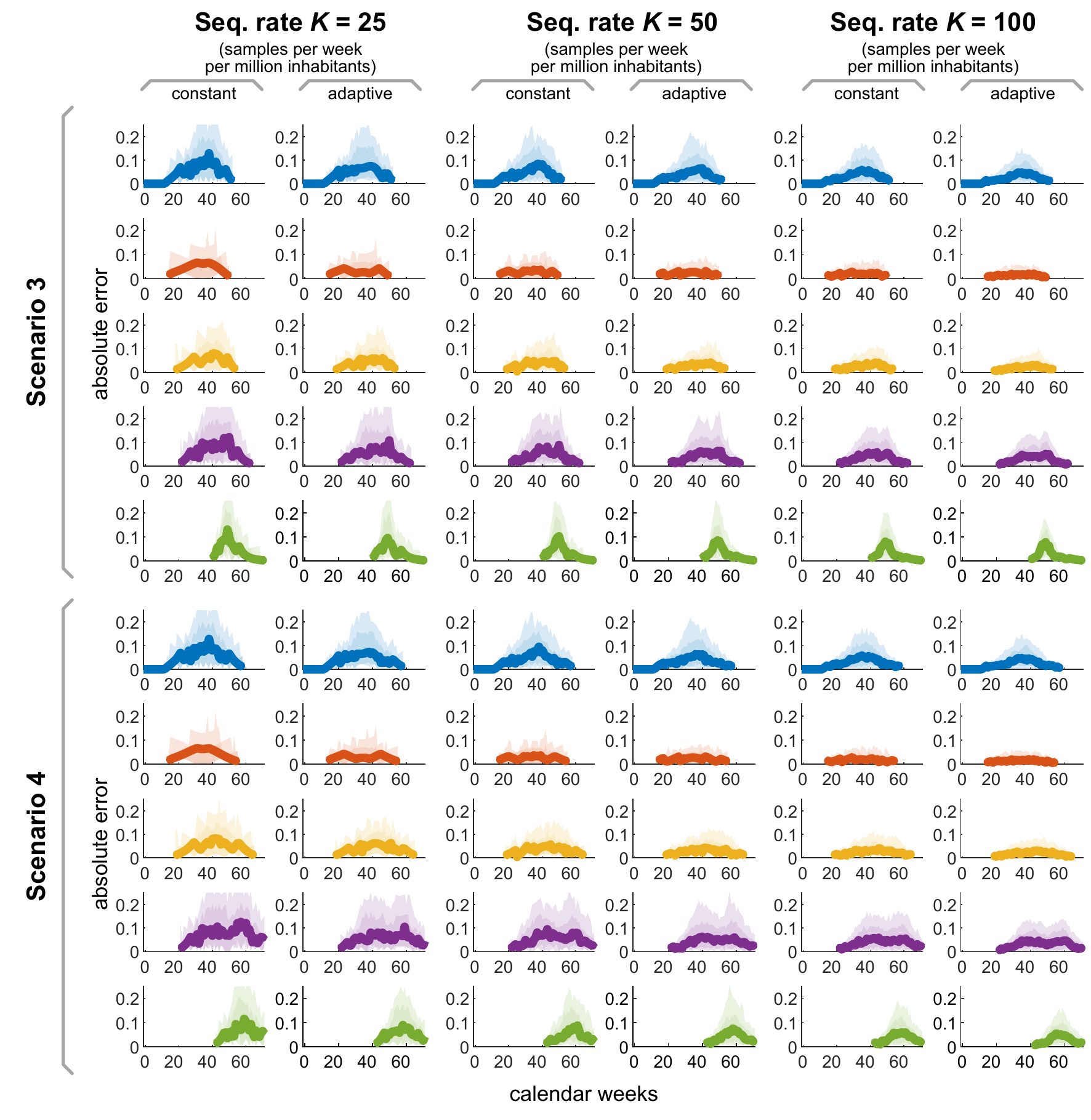}
    \caption{%
        \textbf{Absolute error trends for the variant shares in scenarios 3 and 4.} Solid lines represent the median absolute error between the estimated variant share $\fihat(t)$ and its ground truth \fit. Shaded areas denote sample variability (dark: 68\%, light: 95\%).
        }
    \label{supfig:Results_4}
\end{figure}

Besides substantially reducing the expected detection delay of variants in a population, an adaptive sampling protocol for genomic surveillance can also reduce the mismatch between real and estimated trends for the variant shares (i.e., the fraction of cases that correspond to a given variant). In Figs.~\ref{supfig:Results_3} and~\ref{supfig:Results_4}, we analyse the absolute error trends for the variant shares, i.e., the $|\fit-\fihat(t)|$ trends, for both sampling protocols. 

\end{document}